\documentclass[aps,prc,onecolumn,superscriptaddress,showpacs,showkeys,nofootinbib]{revtex4-2}
\usepackage{graphicx}
\usepackage{bm}
\usepackage{epsfig}
\usepackage{amsfonts}
\usepackage{amssymb,amscd}
\usepackage{subfigure}
\usepackage{xcolor}
\usepackage{comment}
\usepackage{amsmath}
\usepackage{graphicx}
\usepackage{dcolumn}
\usepackage{bm}
\usepackage{hyperref}

\begin{document}

\title{\boldmath Double D meson production in ultraperipheral $pp$, $pPb$ and $PbPb$ collisions}

\author{Ya-Ping Xie}
\email{xieyaping@impcas.ac.cn}
\affiliation{Institute of Modern Physics, Chinese Academy of Sciences,
	Lanzhou 730000, China}
\affiliation{University of Chinese Academy of Sciences, Beijing 100049, China}
\affiliation{State Key Laboratory of Heavy Ion Science and Technology, Institute of Modern Physics,
	Chinese Academy of Sciences, Lanzhou 730000, China}
\author{Victor P. { Gon\c{c}alves}}
\email{barros@ufpel.edu.br}
\affiliation{Institute of Physics and Mathematics, Federal University of Pelotas, Postal Code 354,  96010-900, Pelotas, RS, Brazil}

\begin{abstract}
The production of a $D^+ D^-$ pair by photon - photon interactions at the Large Hadron Collider (LHC)  is investigated considering ultraperipheral proton - proton ($pp$), proton - lead ($pPb$) and lead - lead ($PbPb$) collisions. Assuming that the scattering amplitude for the $\gamma \gamma  \rightarrow D^+ D^-$ process can be described by the Brodsky - Lepage formalism in the heavy - quark approximation, we derive the associated differential distributions and total cross - sections. In particular, in addition to the rapidity and transverse momentum distributions, usually presented in the literature, we also present predictions for the total transverse momentum and  momentum imbalance distributions. Our results indicate that a future experimental analysis of this final state is, in principle, feasible during the high luminosity run of LHC.
\end{abstract}

\maketitle

\section{Introduction}
Two - photon collisions offer a unique possibility to probe the theory of strong interactions~\cite{Terazawa:1973tb,epa,Chernyak:1983ej,Chernyak:2014wra}. 
The simplicity of the initial state and the possibility of studying several different combinations of final states make this process very useful for studying the meson production, its description in terms of distribution amplitudes, and the high energy limit of the QCD dynamics (See, e.g., Refs.~\cite{Ginzburg:1985tp,Ginzburg:1992mi,Qiao:2001wv,Pire:2005ic,Goncalves:2005gv,Goncalves:2006wy,Segond:2007fj,Enberg:2005eq,Carvalho:2015mra,Babiarz:2019sfa,Yang:2020xkl}). 
Such potentiality is one of the motivations for the construction of the future $e^+ e^-$ colliders~\cite{ILC:2013jhg,CEPCStudyGroup:2018ghi,FCC:2018byv,FCC:2018evy,FCC:2025lpp}. However, over the last decades, it became clear that the physics associated with photon - photon interactions can also be investigated in ultraperipheral hadronic collisions at the Large Hadron Collider (LHC). In particular, the double hadron production have been discussed e.g. in Refs.~\cite{Goncalves:2002vq,Goncalves:2003qq,Goncalves:2005rz,Goncalves:2006hu,Klusek:2009yi,Luszczak:2011js,Baranov:2012vu,Klusek-Gawenda:2013dka,Goncalves:2015sfy,Klusek-Gawenda:2017lgt,Andrade:2022rbn,Siddikov:2022bku, Siddikov:2023qbd,Zhang:2024mql,Jia:2025oak,Lebiedowicz:2026qff}.  In ultraperipheral collisions, the  impact parameter larger than the sum of the radius of the incoming hadrons, implying the suppression of the strong interactions, with the hadrons interacting predominantly through its electromagnetic fields, which are sufficiently large to allow the particle production via the photon - photon fusion~\cite{upc,Krauss:1997vr}. 
The final state is very clean, being characterized by the produced system, the intact hadrons and the presence of rapidity gaps.
Additionally, photon - photon interactions in ultraperipheral hadronic collisions have two other advantages: (a) the cross-sections varies as $Z_1^2 Z_2^2 \alpha_{em}^n$, where the $Z_i$ are the number of protons of the incoming hadrons,  rather just as $\alpha_{em}^n$~\footnote{The exponent $n$ depends on the subprocess considered.} in $e^+ e^-$ collisions, and (b) the maximum $\gamma \gamma$ center - of - mass energy that can reached at the LHC is larger than that probed at LEP2. 
As a consequence, LHC can be considered a  bridge to photon - photon collisions at a future $e^+ e^-$ collider.

\begin{figure}[t]
    \centering
        \includegraphics[width=0.48\linewidth]{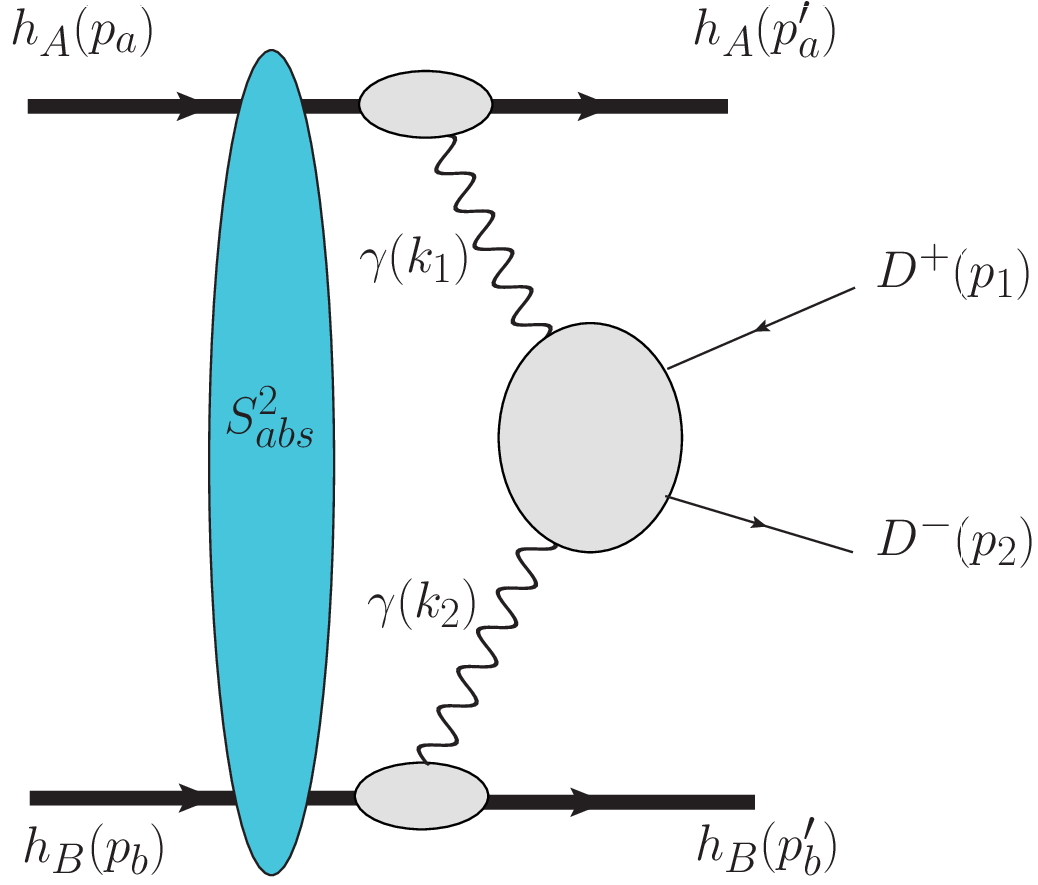}
         \caption{Double $D$ meson production by $\gamma \gamma$ interactions in ultraperipheral hadronic collisions.}
    \label{fig:diagram}
\end{figure}

In our analysis, we will focus on the $D$ meson pair production by $\gamma \gamma$ interactions in ultraperipheral hadronic collisions, represented in Fig.~\ref{fig:diagram}. Such a possibility already was explored  in Ref.~\cite{Luszczak:2011js}, where the cross - sections for ultraperipheral $PbPb$ collisions were estimated using the equivalent photon approximation in the impact parameter space and
assuming distinct approaches to describe the $\gamma \gamma \rightarrow D \bar{D}$ amplitude.
More recently, such analysis was expanded in Ref.\cite{Lebiedowicz:2026qff}  to take into account of the resonant contributions, which are important to low invariant mass of the $D$ meson pair system. 
Distinctly from these previous  studies, here we consider a formalism that, in addition to the impact parameter dependence, takes into account of the transverse momentum and polarization of the emitted photons, discussed e.g. in Refs.~\cite{Krauss:1997vr,Vidovic:1992ik,Wang:2021kxm,Hui:2024nsw}. Such a formalism allow us to estimate in more detail the differential distributions, such as the transverse momentum imbalance of the $D$ meson pair in the final state. Moreover, we will also present, for the first time, the predictions associated with the production of this final state in ultraperipheral $pp$ and $pPb$ collisions.

This paper is organized as follows. In the next Section, we present a brief review of the formalism needed to describe the  exclusive double $D$ meson production by $\gamma \gamma$ interactions in  ultraperipheral hadronic collisions. In particular, we will discuss the main ingredients used in our calculations. In Section \ref{sec:results} we will present our predictions for the differential distributions considering $pp$, $pPb$  and $PbPb$ collisions at the LHC energies. Predictions for the total cross-sections  will also be presented.  Finally, in Section \ref{sec:summary} we will summarize our main conclusions.

\section{Formalism}
The particle production by photon-photon interactions in ultraperipheral collisions is usually estimated using the equivalent photon approximation (EPA)~\cite{epa}, disregarding the photon's transverse momentum and polarization. 
However, in recent years, EPA predictions for dilepton pair production often show deviations from experimental data for the transverse momentum spectrum measured by the STAR Collaboration at RHIC \cite{STAR:2004bzo,STAR:2019wlg,STAR:2018ldd} and the ATLAS and ALICE and  Collaborations at LHC\cite{ATLAS:2018pfw,ATLAS:2020epq,ALICE:2022hvk}. 
Such experimental results have motivated the improvement of theoretical formalism in order to compute polarization-dependent cross-sections as functions of the produced pair's transverse momentum and the impact parameter (See, e.g., Refs.~\cite{Li:2019sin,Klein:2020jom,Xiao:2020ddm,Shao:2022stc,Wang:2021kxm,Klusek-Gawenda:2020eja,Boer:2024cnw,Shi:2024gex}). In what follows, we will consider the approach discussed in detail e.g. in Refs.~\cite{Krauss:1997vr,Vidovic:1992ik,Wang:2021kxm,Hui:2024nsw}, which imply that the cross-section can be expressed as
\begin{align}\label{eq:xsec-def}
\sigma=&\int \frac{\mathrm{d}^2 \boldsymbol{b}_{\perp} \mathrm{d}^3 \boldsymbol{p}_1 \mathrm{~d}^3 \boldsymbol{p}_2}{(2 \pi)^3 2 E_1(2 \pi)^3 2 E_2} S^2_{abs}(\boldsymbol{b}_{\perp})
\Bigg|\int \frac{\mathrm{d}^4 k_1 \mathrm{~d}^4 k_2}{(2 \pi)^4(2 \pi)^4}(2 \pi)^4 \delta^{(4)}\left(k_1+k_2-p_1-p_2\right) 
\mathcal{M}_{\mu \nu}\left(k_1, k_2, p_1, p_2\right) A_1^\mu\left(k_1, b_{\perp}\right) A_2^\nu\left(k_2, 0\right)\Bigg|^2,\nonumber \\
\end{align}
where $\boldsymbol{b}_{\perp}$ is the impact parameter of the collision, defined as the transverse distance between the center of the colliding hadrons, $\boldsymbol{p}_{1,2}$ and $E_{1,2}$ are the three-momenta and energy of the $D^+$ and $D^-$ mesons in the final state, respectively. The photon momenta of the incoming photons are denoted by $k_{1,2}$ and 
$\mathcal{M}_{\mu \nu}$ is the vertex function, which is associated with the $\gamma (k_1) \gamma (k_1) \rightarrow D^+ (p_1) D^- (p_2)$ transition. The classical photon fields in momentum space are~\cite{Krauss:1997vr}
\begin{align}
&A_{1}^{\mu}(k_{1}, b_{\perp})=2\pi Z_1e\frac{F_1(-k_{1}^{2})}{-k_{1}^{2}}\delta(k_{1}\cdot u_{1})u_{1}^{\mu}e^{i\bm{k}_{1\perp}\cdot \bm{b}_{\perp}},\\
&A_{2}^{\mu}(k_{2},0)=2\pi Z_2e\frac{F_2(-k_{2}^{2})}{-k_{2}^{2}}\delta(k_{2}\cdot u_{2})u_{2}^{\mu},
\end{align}
where $u_{1,2}^{\mu} = \gamma_L(1,0,0,\pm v)$ ($\gamma_L$ is the Lorentz factor) and $Z_{1,2}$ are the proton numbers and velocities of the incoming hadrons. An extra phase factor $e^{i\bm{k}_{1\perp}\cdot \bm{b}_{\perp}}$  is included in $A_{1}^{\mu}(k_{1}, b_{\perp})$ to describe the dependence on the impact parameter. Moreover, $F_i(-k_{i}^{2})$ are the associated charge density distributions. Finally, the function $S^2_{abs}(\boldsymbol{b}_{\perp})$ is the absorption factor that assures that we consider only ultraperipheral collisions, which will be assumed to be given by $S^2_{abs}(\boldsymbol{b}_{\perp}) = \Theta(|\boldsymbol{b}_{\perp})| - R_A - R_B)$, with $R_i$ being the hadron radius (For a  detailed discussion about the absorption factor, see e.g. Refs~\cite{Klusek-Gawenda:2010vqb,Azevedo:2019fyz}). Performing the substitutions and scalar products, the cross-section can be expressed by
\begin{align}
\sigma = &\frac{Z_1^2 Z_2^2 e^{4}}{(4vk^0_1k^0_2)^2}\int\frac{\mathrm{d}^{2}\boldsymbol{b}_{\perp}\mathrm{d}^{3}\boldsymbol{p}_{1}\mathrm{d}^{3}\boldsymbol{p}_{2}}{(2\pi)^{2}E_{1}E_{2}} S^2_{abs}(\boldsymbol{b}_{\perp})  \\
&\hspace{-0.5cm}\times \biggl|\int\frac{\mathrm{d}^{2}\boldsymbol{k}_{1\perp}\mathrm{d}^{2}\boldsymbol{k}_{2\perp}}{(2\pi)^{2}(2\pi)^{2}}\delta^{(2)}(\boldsymbol{k}_{1\perp}+\boldsymbol{k}_{2\perp}-\boldsymbol{p}_{1\perp}-\boldsymbol{p}_{2\perp})
&\hspace{-0.5cm}\times\mathcal{M}^{ij}(k_{1},k_{2},p_{1},p_{2})e^{i\bm k_{1\perp}\cdot \bm b_{\perp}}k_{1\perp,i}k_{2\perp,j}\frac{F_1(-k_{1}^{2})}{-k_{1}^{2}}\frac{F_2(-k_{2}^{2})}{-k_{2}^{2}}\biggr|^{2},\notag 
\end{align}
where $k_{1\perp}$ and $k_{2\perp}$ represent the transverse momentum of the incoming photons.


In order to derive theoretical predictions for the differential distributions that are measured by the experimental collaborations, we will express the cross - section in terms of the rapidities $y_{1,2}$ of the $D$ mesons and the  variables $\boldsymbol{q}_\perp$ and $\boldsymbol{P}_\perp$, defined by
\begin{eqnarray}
\boldsymbol{q}_\perp = {\boldsymbol{p}_{1}+\boldsymbol{p}_{2}} \,\,\,\,\mbox{and} \,\,\,\,\boldsymbol{P}_\perp = \frac{\boldsymbol{p}_{1}-\boldsymbol{p}_{2}}{2} \,\,
\end{eqnarray}
which represent the sum and difference of the outgoing particle transverse momenta, respectively. The cross-section will be given by
\begin{widetext}
	\begin{align}\label{eq:dis}
	&\frac{\mathrm{d}\sigma}{\mathrm{d}^{2}\boldsymbol{b}_{\perp}\mathrm{d}^{2}\boldsymbol{P}_\perp\mathrm{d}^{2}\boldsymbol{q}_\perp\mathrm{d}y_1\mathrm{d}y_2} = \frac{Z_1^2 Z_2^2\alpha_{em}^2}{m^{4}}S^2_{abs}(\boldsymbol{b}_{\perp})\int\!\frac{[\mathrm{d}\mathcal{K}_{\perp}]}{(2\pi)^8}\!\!\sum_{s_{1},s_{2}} \big[\!\left(\mathcal{M}_{-+}\mathcal{M}_{+-}^{*}\!+\!\mathcal{M}_{+-}\mathcal{M}_{-+}^{*}\right)\cos(\phi_{\boldsymbol{k}_{1\perp}}\!\!+\!\phi_{\boldsymbol{k}_{2\perp}}\!\!-\!\phi_{\bar{\boldsymbol{k}}_{1\perp}}\!\!-\!\phi_{\bar{\boldsymbol{k}}_{2\perp}}) \nonumber \\
	&+ \!\left(\mathcal{M}_{--}\mathcal{M}_{--}^{*}\!+\!\mathcal{M}_{++}\mathcal{M}_{++}^{*}\right)\cos(\phi_{\boldsymbol{k}_{1\perp}}\!\!-\!\phi_{\boldsymbol{k}_{2\perp}}\!\!-\!\phi_{\bar{\boldsymbol{k}}_{1\perp}}\!\!+\!\phi_{\bar{\boldsymbol{k}}_{2\perp}}\!)\!+\! \left(\mathcal{M}_{--}\mathcal{M}_{++}^{*}\!+\!\mathcal{M}_{++}\mathcal{M}_{--}^{*}\right)\cos(\phi_{\boldsymbol{k}_{1\perp}}\!\!-\!\phi_{\boldsymbol{k}_{2\perp}}\!\!+\!\phi_{\bar{\boldsymbol{k}}_{1\perp}}\!\!-\!\phi_{\bar{\boldsymbol{k}}_{2\perp}}\!)\big],
	\end{align}
\end{widetext}
where we only keep the azimuthal independent terms, averaging over $\phi_{\boldsymbol{P}_\perp}$ and $\phi_{\boldsymbol{q}_\perp}$. Moreover, $m$ is the invariant mass of the final state and we have assumed a shorthanded notation, where  $\mathcal{M}_{\lambda_1 \lambda_2} \equiv \mathcal{M}_{s_1 s_2,\lambda_1 \lambda_2}|_{\phi_{\boldsymbol{p}_{1\perp}}=\phi_{\boldsymbol{p}_{2\perp}}-\pi=0}$ and
\begin{align}
{\cal \int} [\mathrm{d}{\cal K}_\perp] \equiv & \int \mathrm{d}^{2}\bm k_{1\perp}\mathrm{d}^{2}\bm k_{2\perp}\mathrm{d}^{2}\bar {\bm k}_{1\perp}\mathrm{d}^{2}\bar {\bm k}_{2\perp}e^{i(\bm k_{1\perp}-\bar {\bm k}_{1\perp})\cdot \bm b_{\perp}} \times \delta^{(2)}(\bm k_{1\perp}+\bm k_{2\perp}-\bm q_{\perp}) \delta^{(2)}(\bar {\bm k}_{1\perp}+\bar {\bm k}_{2\perp}-{\bm q}_{\perp}) \notag \\
&\times \mathcal{F}_1(x_1,{\bm k}_{1\perp}^{2})\mathcal{F}_2(x_2, {\bm k}_{2\perp}^{2})\mathcal{F}_1(x_1, \bar {\bm k}_{1\perp}^{2})\mathcal{F}_2(x_2, \bar {\bm k}_{2\perp}^{2}).
\end{align}
We have that the cross-section is now expressed in terms of the helicity amplitudes $\mathcal{M}_{\lambda_1 \lambda_2}$ and the functions $\mathcal{F}_i(x_i, {\bm k}_{i\perp}^{2})$, which allow us to estimate the probability of finding a photon with a longitudinal momentum $x_i$ and transverse momentum ${\bm k}_{i\perp}$ (See, e.g., Refs.~\cite{Wang:2021kxm,Klusek-Gawenda:2020eja,Boer:2024cnw,Shi:2024gex}), and that are expressed by
\begin{align}
{\cal F}(x_i, \bm{k}_{i,\perp}^2) = |\bm k_{i\perp}|\frac{F(\bm{k}_{i\perp}^2 + x_i^2 m_p^2)}{\bm{k}_{i\perp}^2 + x_i^2 m_p^2}.
\end{align}
The  photon momentum fractions of the incoming photons in the nucleon-nucleon c.m. frame  are given by
\begin{align}\label{eq:energy_frac}
x_{1,2} &= \frac{\sqrt{P_{\perp}^2 + m_D^2}}{\sqrt{s_{N\!N}}}\left(e^{\pm y_{1}} + e^{\pm y_{2}}\right),
\end{align}
where $m_D$ is the $D$ meson mass and $\sqrt{s_{N\!N}}$ is the center - of - mass energy.
In our analysis,  we will assume  the nuclear form factor obtained by a numerical approximation to the Woods-Saxon potential \cite{Klein:2016yzr} that is extensively adopted in the literature, which is described by
\begin{align}\label{eq:nucharge}
F\left(|k|\right)=\frac{3\left[\sin \left(|k| R_A\right)-|k| R_A \cos \left(|k|R_A\right)\right]}{\left(|k| R_A\right)^3\left(a^2 |k|^2+1\right)},
\end{align}
with $a=0.7$ fm and $R_{A}=1.1A^{1/3}$ fm. 
On the other hand, for the proton, we will assume the dipole form factor
\begin{eqnarray}
F(|k|) = \frac{1}{(1+|k|^2/0.71)^2}\,\,.
\end{eqnarray}

The last ingredient needed to calculate the double $D$ meson production cross-section is the amplitude for the $\gamma \gamma \rightarrow D^+ D^-$ process. We will focus on the  production of a $D$ meson pair with large invariant mass ($m > 4.0 $ GeV), where a perturbative approach is justified and the resonant contributions are expected to be negligible. In our analysis, following Ref.~\cite{Baek:1994kj}, we will assume Brodsky - Lepage formalism~\cite{Lepage:1979za,Lepage:1980fj,Brodsky:1981rp} in the heavy - quark approximation, which implies that    
\begin{eqnarray}
M_{\lambda \lambda'} &=&
\frac{2F^{\gamma \gamma}} {(1-z^2)^2}\Bigg\{
(1+z^2)\Big [e_Q^2 F_{\lambda \lambda'}
+e_q^2 \widetilde{F}_{\lambda \lambda'}\Big ]
-2 e_Q e_q G_{\lambda \lambda'}
-(1-z^2)\Big [\frac{1-x}{x} e_Q^2 H_{\lambda \lambda'}
+ \frac {x}{1-x} e_q^2 \widetilde{H}_{\lambda \lambda'}\Big ] \Bigg\},
\end{eqnarray}
where
\begin{eqnarray}
F^{\gamma \gamma} & = & 
\frac{16\pi^2\alpha_{em}\alpha_sC_F}{3\hat {s}^2x^2(1-x)^2}
\left[ \frac{f_D}{m_D} \right]^2, \\
F_{\lambda \lambda'} & = &
\Big [(1-x)[2-x(\hat s +2)]+ \frac{\hat s}{2}\delta_{\lambda,-\lambda'}\Big ](\beta^2-z^2), \\
\widetilde{F}_{\lambda \lambda'} & = &
\Big [x[2-(1-x)(\hat s +2)]+\frac{\hat s}{2}\delta_{\lambda,-\lambda'}\Big ](\beta^2-z^2), \\
G_{\lambda \lambda'}& = &
\Big [\frac{\hat s}{2}[1+z^2+(1-z^2)\delta_{\lambda, \lambda'}]-2x(1-x)(2+\hat s)\Big ]
(\beta^2-z^2)-[\hat s(1-z^2)-2(3-z^2)]\delta_{\lambda,\lambda'}, \\
H_{\lambda \lambda'} & = &
[2-x(\hat s +2)]\Big [(1-x)(\beta^2-z^2)+z^2\delta_{\lambda,\lambda'}\Big ]+x(\hat s+2)
\delta_{\lambda,\lambda'}, \\
\widetilde{H}_{\lambda \lambda'}& = &
\Big [2-(1-x)(\hat s +2)\Big ]\Big [x(\beta^2-z^2)+z^2\delta_{\lambda,\lambda'}\Big]
+(1-x)(\hat s+2)
\delta_{\lambda, \lambda'},
\end{eqnarray}
with $z = \beta \cos\theta$, $\theta$ being the scattering angle between the photon and the heavy meson, 
$\beta= \sqrt{ 1-{4}/{\hat s}}$ and $\hat s \equiv {W_{\gamma\gamma}^2}/{m_D^2}$, where $W_{\gamma\gamma}$ is photon - photon center-of-mass energy. Moreover, $C_F=\frac{4}{3}$, $x = (m_D - m_c)/m_D$, with $m_c$ being the charm mass, and  $f_D$ is the meson decay constant. It is important to emphasize that in Ref.~\cite{Luszczak:2011js}, the predictions associated with the heavy quark approximation have been compared with those derived assuming a distinct approach to estimate the distribution amplitude and obtained similar results for the $D^+D^-$ production.

\begin{table}[t]
	\begin{tabular}{|c|c|c|}
\hline
\hline
{\bf Colliding system} & {\bf Center-of-mass energy} & {\bf Total cross-section} \\
\hline
$pp$  & $\sqrt{s} = 14$ TeV &   0.15 pb \\
\hline
$pPb$ & $\sqrt{s} = 8.1$ TeV & 0.50 nb \\
\hline
$PbPb$ & $\sqrt{s} = 5.02$ TeV & 1.3 $\mu$b \\
	\hline
	\end{tabular}
	\caption{Predictions for the total cross - sections associated with the   production of a $D^+D^-$ pair by $\gamma \gamma$ interactions in ultraperipheral $pp$, $pPb$ and $PbPb$ collisions at the LHC energies.}
	\label{Tab:cs}
\end{table}

\section{Results}
\label{sec:results}
In what follows we will present our predictions for the production of a $D^+D^-$ pair in ultraperipheral hadronic collisions at the LHC, derived using the formalism described in the previous section. We will consider $pp /\,  pPb /\, PbPb$  collisions at  $14/\,8.1/\,5.02$ TeV. Initially, in Table \ref{Tab:cs}, we present our results for the total cross-sections. Due to the $Z^2$ dependence of the photon spectra, we have that the following hierarchy is approximately valid for the $D^=D^-$ production induced by $\gamma \gamma$ interactions: $\sigma_{PbPb} \approx Z^2 \cdot \sigma_{pPb} \approx Z^4 \cdot \sigma_{pp}$, with $Z = 82$. The predictions for $pp$ and $pPb$ collisions are presented here for the first time. For $PbPb$ collisions, our results agree with those derived in Ref.~\cite{Luszczak:2011js}. Considering that the expected integrated luminosities for the high luminosity run of the LHC are 3000 fb$^{-1}$ / 13 nb$^{-1}$ 
for $pp / Pb Pb$ collisions~\cite{Apollinari:2017lan}, we predict  that the
number of events per year in $pp$ ($PbPb$) collisions will be $\approx 450000 \, (16900)$ events. It is important to emphasize that these numbers must be considered as a lower bound, since the resonant contributions~\cite{Lebiedowicz:2026qff}, not taken into account in our analysis, are expected to dominate at small values of the invariant mass of the $D^+ D^-$ system.

\begin{figure}[t]
    \centering
        \includegraphics[angle=-90, width=0.43\linewidth]{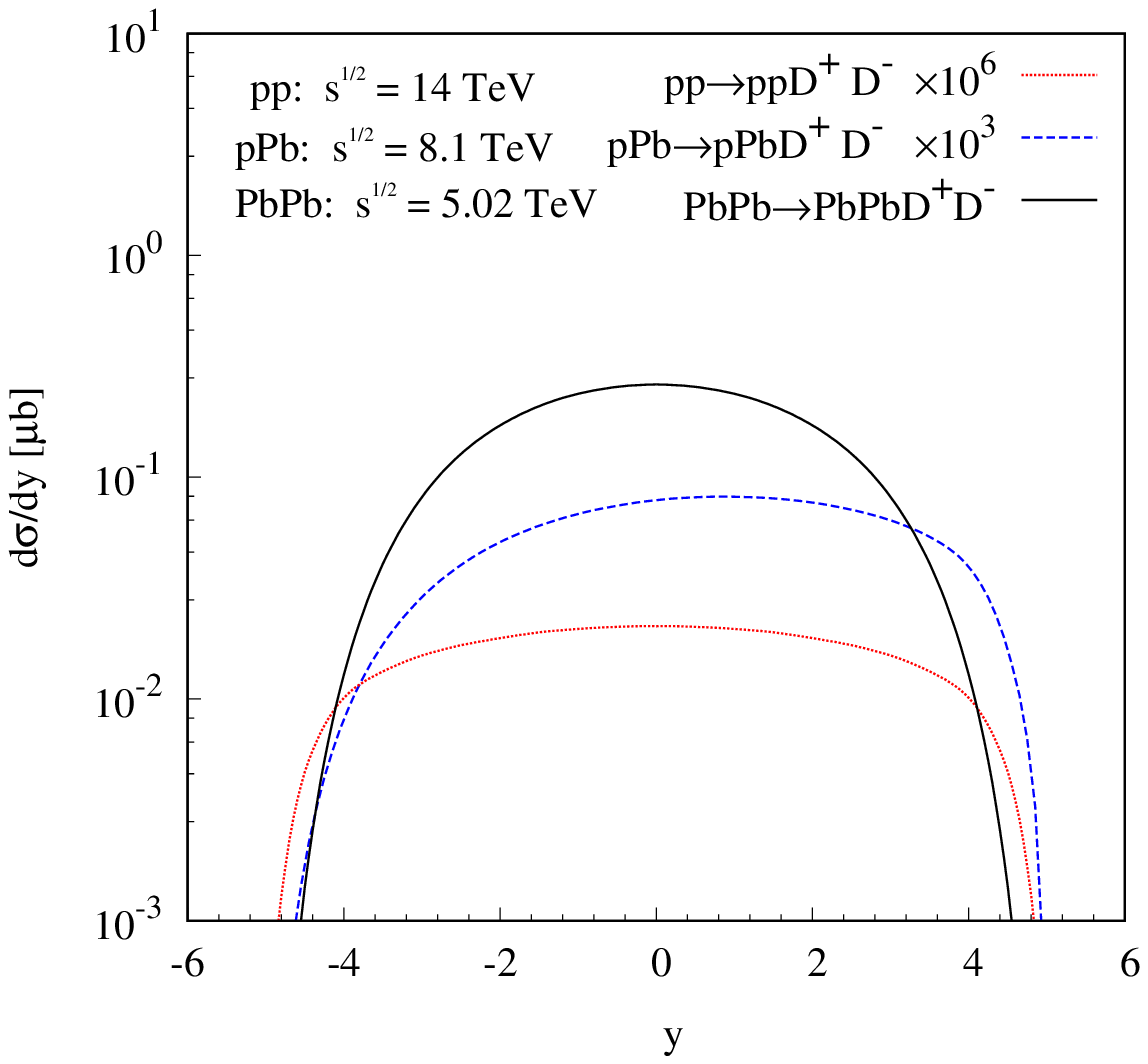}
     \includegraphics[angle=-90,width=0.43\linewidth]{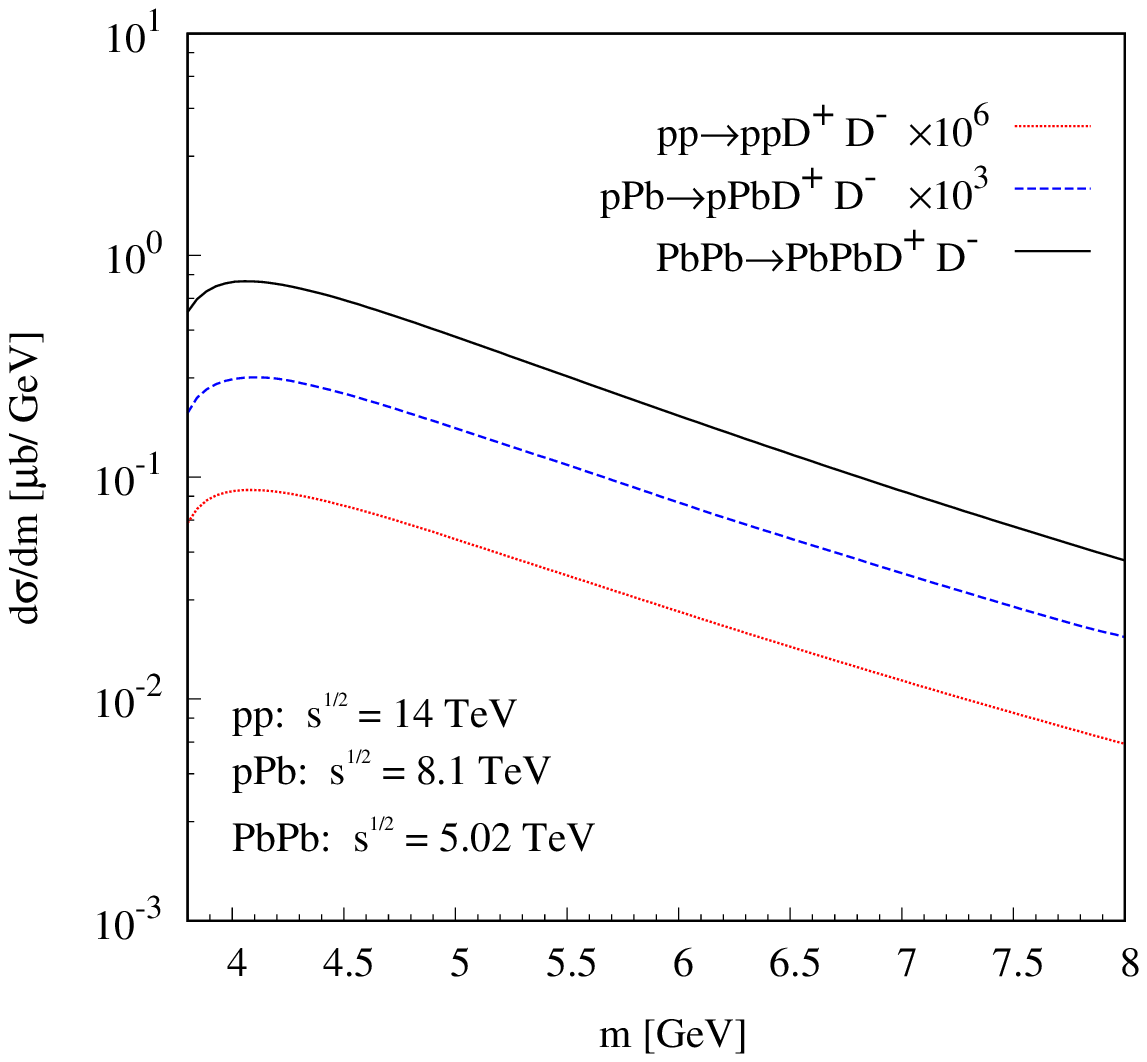} 
         \caption{Predictions for the rapidity (left panel) and invariant mass (right panel) distributions for the production of a $D^+ D^-$ pair in ultraperipheral $pp$, $pPb$ and $PbPb$ collisions at the LHC energies.}
    \label{fig:dist1}
\end{figure}

\begin{figure}[t]
    \centering
          \includegraphics[angle=-90,width=0.48\linewidth]{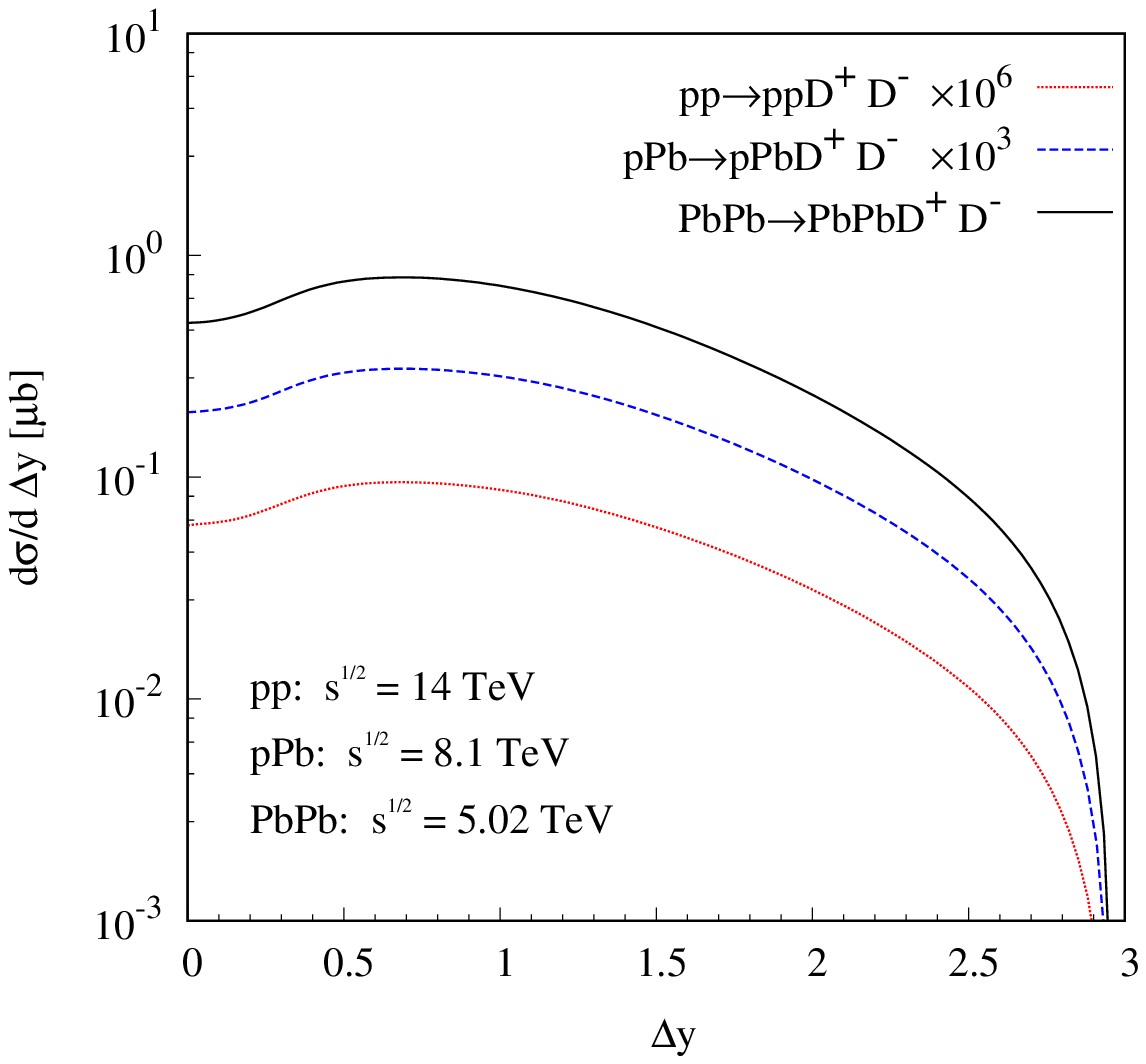} 
    \caption{Predictions for the rapidity difference distribution derived considering $pp$, $pPb$ and $PbPb$ collisions.}
    \label{fig:dist3}
\end{figure}

In Fig.~\ref{fig:dist1} we present our results for the rapidity and invariant mass  distributions for the production of a $D^+ D^-$ pair in ultraperipheral $pp$, $pPb$ and $PbPb$ collisions at the LHC energies. The predictions for $pp$ and $pPb$ collisions have been rescaled by a constant factor in order to be presented in the same plot.
For the rapidity distributions (left panel), due to the asymmetry in the proton and nuclear photon fluxes present in the initial state, we predict an asymmetric distribution in $pPb$ collisions. For $pp$ and $PbPb$ collisions, we predict symmetric distributions, which differ in normalization. In particular, for $PbPb$ collisions, the distribution is similar to that presented in Ref.~\cite{Luszczak:2011js}  for the heavy - quark approximation. For the invariant mass distributions (right panel), we predict similar distributions for $pp$, $pPb$ and $PbPb$ collisions, differing only in normalization. Such a result is expected, since the behavior of the distribution is  determined by the $\gamma \gamma \rightarrow D^+ D^-$ subprocess. Finally, in Fig.~\ref{fig:dist3}, we present our predictions for the differential distribution as a function of the rapidity difference $\Delta y \equiv |y_1 - y_2|$ between the mesons. In this case, we predict a maximum for 
$\Delta y \approx 0.5$, decreasing for larger values of $\Delta y$.

\begin{figure}[t]
    \centering
               \includegraphics[angle=-90,width=0.43\linewidth]{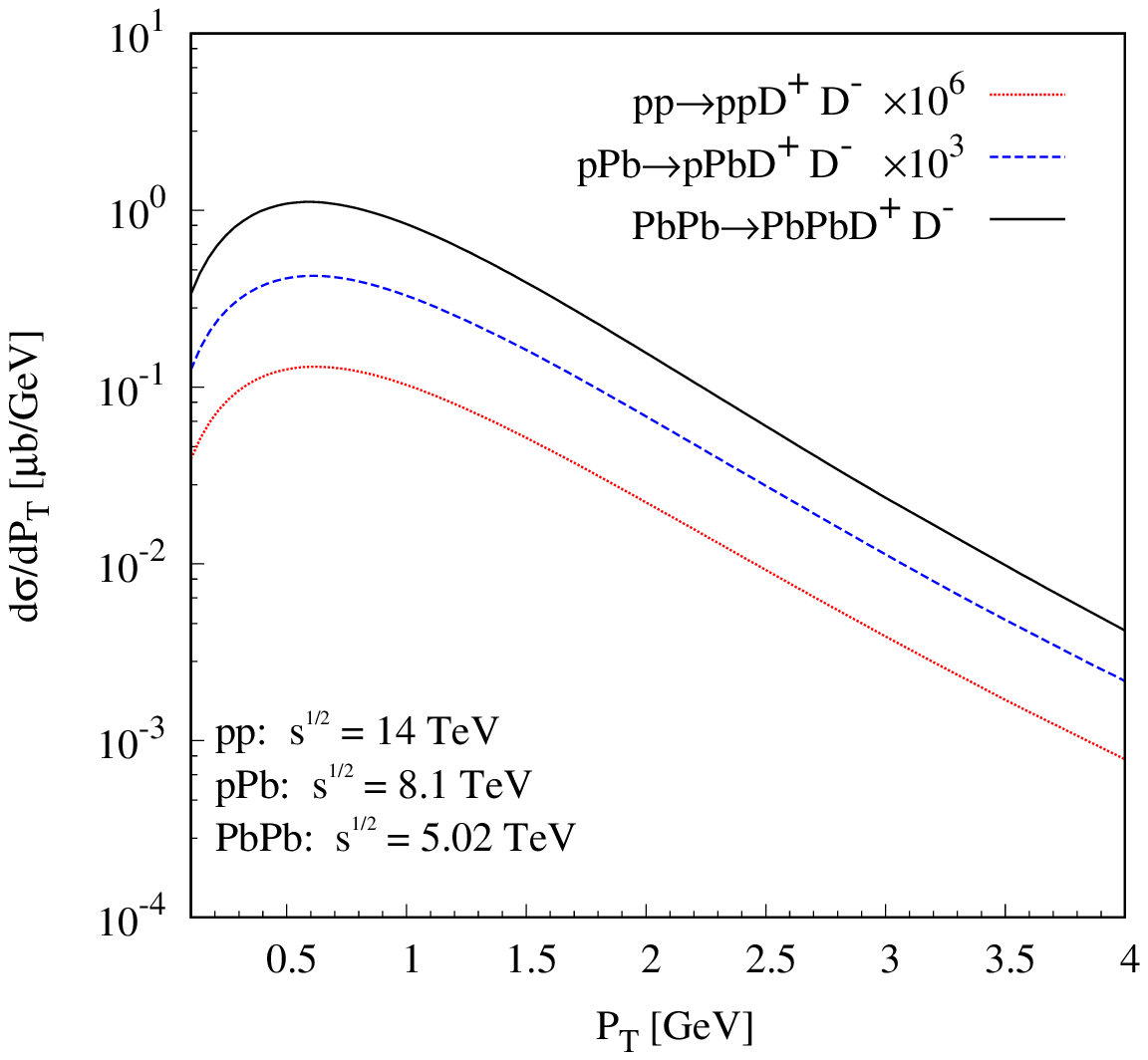}
          \includegraphics[angle=-90,width=0.43\linewidth]{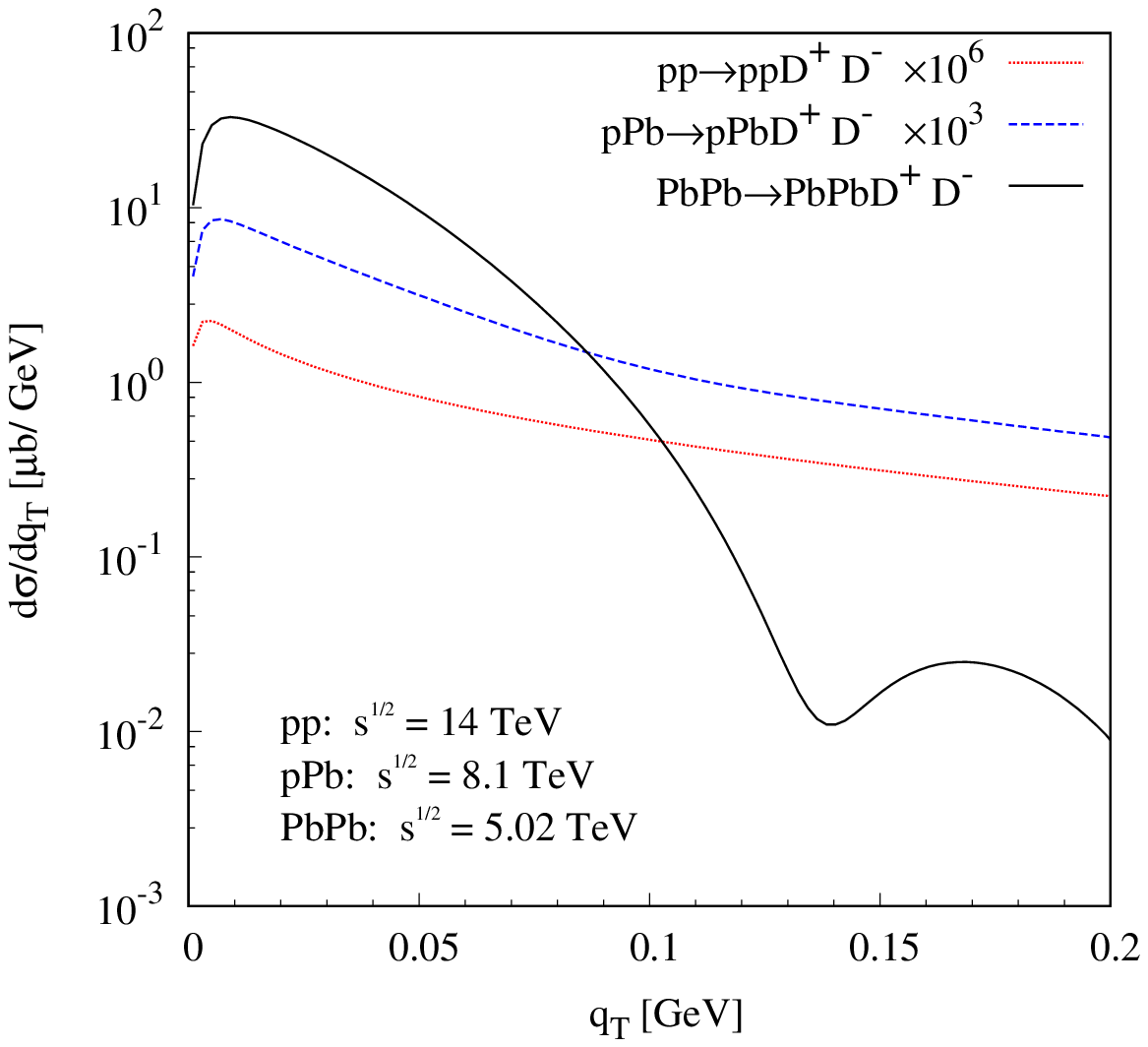}
    \caption{Predictions for the total transverse momentum  of the $D^+ D^-$  pair (left panel) and transverse momentum imbalance (right panel)  differential distributions derived considering $pp$, $pPb$ and $PbPb$ collisions.}
    \label{fig:dist2}
\end{figure}

One of the advantages of the formalism used in this paper, is that it allow us to estimate the differential distributions  as a function of the total transverse momentum of the $D^+ D^-$  pair 
$\boldsymbol{P}_\perp$ and the transverse momentum imbalance $\boldsymbol{q}_\perp$. We have that if the transverse momentum of the photons are disregarded, as usually assumed in the literature, the $D$ mesons in the final state will be characterized by ${\boldsymbol{p}_{1} = - \boldsymbol{p}_{2}}$. Therefore, the analysis of the $\boldsymbol{q}_\perp$ distribution is important to improve the  description of transverse momenta of the incoming photons and constrain the photon Wigner distribution~\cite{Klusek-Gawenda:2020eja,Boer:2024cnw,Shi:2024gex}.
The associated distributions are presented in Fig.~\ref{fig:dist2}. For the $\boldsymbol{P}_\perp$ distributions, we predict similar shapes for $pp$, $pPb$ and $PbPb$ collisions, with the results differing only in magnitude and a peak for $|\boldsymbol{P}_\perp| \approx 0.5$ GeV. In contrast, for the $\boldsymbol{q}_\perp$ distribution, the shape and the position of the peak depend on the colliding system. In particular, for $PbPb$ collisions, we predict the presence of a dip in the distribution, which is directly associated with the transverse momentum dependence of the nuclear form factor.

\section{Summary}
\label{sec:summary}
During the last decades, the possibility of probing the hadron production by two - photon fusion became a reality in ultraperipheral hadronic collisions, which motivate the investigation of distinct final states. One of the most promissing is the double hadron production, which can probe the distribution amplitudes, as well as the QCD dynamics at high energies. 
In this paper we have investigated the double $D$ meson production by $\gamma \gamma$ interactions in ultraperipheral hadronic collisions at the LHC energies considering a theoretical formalism that takes into account the tranverse momentum of the initial photons, in addition to the impact parameter dependence of the collision. We have estimated the total cross-sections and associated differential distributions for the production of a $D^+D^-$ pair in $pp$, $pPb$ and $PbPb$ collisions. Our results indicated that, in principle, a future experimental analysis of this final state will be feasible in the high luminosity run of LHC, which will allow us to improve our undestanding of hadron production by $\gamma \gamma$ interactions and probe the photon Wigner distribution.

\section*{Acknowledgments}
 Y. P. Xie gives many thanks to Cheng Zhang for the useful discussions.
  V.P.G. was partially supported by CNPq,  FAPERGS and INCT-FNA (Process No. 408419/2024-5).
This work is partially supported by the NFSC grant (Grant No. 12293061) and National Key R\&D Program of China (Grant No. 2024YFA1611000)

\end{document}